\documentstyle[pra,aps,manuscript,epsfig]{revtex}
\begin{document}
\title{Multidimensional solitons in periodic potentials}
\author{Bakhtiyor B. Baizakov$^1$, Boris A. Malomed$^2$, and Mario Salerno$^1$}
\address{$^1$ Dipartimento di Fisica "E.R. Caianiello"\\
and Istituto Nazionale di Fisica della Materia (INFM), \\
Universit\'a di Salerno, I-84081 Baronissi (SA), Italy, \\
$^2$ Department of Interdisciplinary Studies, Faculty of Engineering, \\
Tel Aviv University, Tel Aviv 69978, Israel}
\date{\today}
\maketitle

\begin{abstract}
The existence of stable solitons in two- and three-dimensional (2D and 3D)
media governed by the self-focusing cubic nonlinear Schr\"{o}dinger equation
with a periodic potential is demonstrated by means of the
variational approximation (VA) and in direct simulations. The potential
stabilizes the solitons against collapse. Direct physical realizations are a
Bose-Einstein condensate (BEC) trapped in an optical lattice, and
a light beam in a bulk Kerr medium of a photonic-crystal type. In the
2D case, the creation of the soliton in a weak lattice potential is possible
if the norm of the field (number of atoms in BEC, or optical power in the
Kerr medium) exceeds a threshold value (which is smaller than the critical
norm leading to collapse). Both ``single-cell" and ``multi-cell" solitons
are found, which occupy, respectively, one or several cells of the periodic
potential, depending on the soliton's norm. Solitons of the former type and
their stability are well predicted by VA. Stable 2D vortex solitons are
found too.
\end{abstract}
\pacs{PACS: 03.75.Lm, 03.75.kk, 05.45.Yv}

Recent successful observation \cite{khaykovich,strecker} and theoretical
consideration \cite{Lincoln,khawaja} of one-dimensional (1D) solitons in
self-attractive Bose-Einstein condensates (BECs) with a negative scattering
length raises a question whether higher-dimensional solitons can be created
in BECs \cite{salasnich}. One possibility, recently proposed in Refs. \cite
{abdullaev,saito}, is to induce self-trapping of a 2D condensate by means of
ac external magnetic field via the Feshbach resonance, which makes the
scattering length a periodically sign-changing one (a similar mechanism was
earlier shown to stabilize light beams in a bulk optical medium consisting
of periodically alternating self-focusing and defocusing layers
\cite{Isaac}). On the other hand, an efficient technique to control BEC dynamics is
based on optical lattices (OLs), i.e., interference patterns generated by
laser beams illuminating the condensate (see earlier works \cite{OLearlier}
and recent ones \cite{OLlater}). A possibility to create higher-dimensional
gap solitons in \emph{self-repulsive} BECs loaded into 2D or 3D OL was
proposed in Ref. \cite{baizakov}, where it was shown that soliton-like
structures emerge due to modulational instability of BEC in the periodic
potential (a similar mechanism for the 1D case was investigated in Refs.
\cite{KS02,1DgapSolitons}). Nonlinear localization of atomic Bloch waves
in the form of 2D matter-wave gap solitons was shown by numerical solution
of the corresponding eigenvalue problem \cite{ostrovskaya}.

The objective of the present work is to show that \emph{stable} 2D and 3D
solitons can be created in a more straightforward (and more relevant to the
experiment) way, namely, in self-attractive BECs trapped in OLs. In the
absence of the external potential, stationary soliton solutions to the
corresponding self-focusing nonlinear Schr\"{o}dinger (NLS) equation can be
easily found, but they are unstable due to the, respectively, weak and
strong collapse in the 2D and 3D NLS equation \cite{Berge}. We demonstrate
that periodic potentials of the OL type can stabilize 2D and 3D solitons of
various types (in particular, both fundamental and vortex solitons are found
to be stable in the 2D case).

The same model finds another physical realization in the 2D case, in terms
of a light beam propagating in a bulk self-focusing Kerr medium with
transverse modulation of the refractive index, i.e., a medium of a
photonic-crystal (PhC) type (see, e.g., Ref. \cite{PhotCryst}). However, in
contrast to recently considered nonlinear PhC models, which assume
transverse modulation of the Kerr coefficient, we consider the case when
only the refractive index is modulated, while the nonlinearity is uniform.
Soliton dynamics in a 1D version of this optical system, i.e., as a matter
of fact, a multichannel nonlinear planar waveguide, was studied earlier in
Ref. \cite{Wang} (after the submission of the original version of this
paper, a preprint has appeared which considers a very similar PhC model for
the 2D case \cite{our_friends}).

The present model is based on the rescaled Gross-Pitaevskii (GP) equation
\cite{review} for the wave function $u(\mathbf{r},t)$,
\begin{equation}
iu_{t}+\left[ \nabla ^{2}+V(\mathbf{r})+|u|^{2}\right] u=0  \label{gpe}
\end{equation}
(in optical models, it is the NLS equation, $t$ being the propagation
distance). Here, $V(\mathbf{r})$ is the external potential, which contains
the parabolic-trap and OL terms,
\begin{equation}
V(\mathbf{r})=\left( \omega ^{2}/2\right) r^{2}+\varepsilon \lbrack \cos
(2x)+\cos (2y)+\cos (2z)], \quad \mathbf{r}\equiv \left\{ x,y,z\right\} ,
\label{potential}
\end{equation}
i.e., the OL period is normalized to be $\pi $ (in the optical model, the
parabolic trap is absent). The only dynamical invariant of Eq. (\ref{gpe})
is the norm (number of atoms in BEC, or the beam's power in the optical
model), $N=\int |u(\mathbf{r})|^{2}d\mathbf{r}$.

Analytical predictions for solitons can be obtained by means of the
variational approximation (VA), which was developed in nonlinear optics (see
a review \cite{malomed}), and successfully applied to BECs too \cite
{Lincoln,abdullaev,BECvariational}, including BECs trapped in OL \cite
{Smerzi} (VA in the 1D version of this model was earlier elaborated for the
above-mentioned multichannel nonlinear optical waveguide \cite{Wang}).
Following these works, we adopt the Gaussian ansatz for fundamental
solitons,
\begin{equation}
u(\mathbf{r},t)=A\exp \left( -i\mu t-ar^{2}/2\right) ,  \label{ansatz}
\end{equation}
with an arbitrary negative constant $\mu $ and positive ones $a$ and $A$ (we
set $\varepsilon >0$ in Eq. (\ref{potential}), hence the central point of
the ansatz must be chosen at the potential minimum, $\mathbf{r}=0$). In the
case of BEC, $\mu $ is the chemical potential, and in optics $-\mu $ is the
propagation constant. Substituting the ansatz into the Lagrangian $L$
corresponding to Eq. (\ref{gpe}), we perform spatial integration, and derive
equations for $a$ and $A$ in the form $\partial L/\partial a=\partial
L/\partial A=0$ ($\mu $ is not a variational degree of freedom, but rather
an intrinsic parameter of the soliton family).

Below, we display variational results for $\omega =0$ in the GP/NLS equation
(\ref{gpe}), which implies that the soliton has a size much smaller than
that imposed by the parabolic trap. In the 2D case, the variational
procedure amounts to an equation which determines $a$ in terms of the norm,
$N=\pi A^{2}/a$ [this expression corresponds to the ansatz (\ref{ansatz})],
and another equation that yields the chemical potential/propagation
constant:
\begin{equation}
N=4\pi \left( 1-2\varepsilon a^{-2}e^{-1/a}\right) ,\quad \mu =2\varepsilon
\left( 2a^{-1}-1\right) e^{-1/a}-a.  \label{2Dvar}
\end{equation}
Note that, if $\varepsilon =0$, Eqs. (\ref{2Dvar}) yield $N=4\pi $. In the
present notation, it is nothing else but a known variational prediction for
the critical norm in the 2D NLS equation, which simultaneously is a
universal value of the norm corresponding to the unstable 2D soliton. With
$\varepsilon \neq 0$, Eq.(\ref{2Dvar}) shows that the number of particles
attains a minimum (threshold) value,
\begin{equation}
N_{\mathrm{thr}}=4\pi \left( 1-8e^{-2}\varepsilon \right) ,  \label{thr}
\end{equation}
at $a=1/2$. This means that there is a threshold value of the norm which is
necessary to create the 2D soliton; however, the threshold really exists
only if it is positive. The expression (\ref{thr}) shows that the threshold
exists for a relatively weak lattice, and it disappears if $\varepsilon $
exceeds the value $\varepsilon _{0}=e^{2}/8\approx \allowbreak
0.\,\allowbreak 92$.

It follows from Eq. (\ref{2Dvar}) that the norm $N$ cannot exceed the
above-mentioned critical value, $N_{\mathrm{cr}}=4\pi $, which means that VA
predicts a family of fundamental 2D solitons in the interval
$N_{\mathrm{thr}}<N<N_{\mathrm{cr}}$ (recall
$N_{\mathrm{thr}}\equiv 0$ if $\varepsilon
>\varepsilon _{0}$). VA makes it also possible to predict stability of the
solitons, on the basis of the \textit{Vakhitov-Kolokolov} (VK) criterion
\cite{Berge}, which applies if the solution to Eqs. (\ref{2Dvar}) is
obtained in a form $\mu =\mu (N)$: the necessary stability condition is
$d\mu /dN<0$.

A numerical solution of Eqs. (\ref{2Dvar}), produces two different branches
of the solution, see an example in Fig. 1a. An analytical consideration of
the near-critical case, $1-N/4\pi \rightarrow +0$, also yields two different
solutions:
\begin{equation}
\mu _{1}(N)\approx -\left( \sqrt{\frac{2\varepsilon }{1-N/4\pi }}
+2\varepsilon \right) , \quad \mu _{2}(N)\approx \left[ \ln \left( 1-N/4\pi
\right) -\ln \left( 2\varepsilon \right) \right] ^{-1}\,.  \label{mu}
\end{equation}
Both the numerical and analytical solutions demonstrate that one branch
[$\mu _{1}(N)$ in Eq. (\ref{mu})] satisfies the VK criterion, and the other
one [$\mu _{2}(N)$ in Eq. (\ref{mu})] does not. Direct simulations (see
inset to Fig. 1a and more results below) confirm that the variational ansatz
satisfying the VK criterion gives rise to a stable soliton, whose form is
quite close to the predicted one.

Qualitatively, the appearance of the family of stable solitons in the
interval $N_{\mathrm{thr}}<N<N_{\mathrm{cr}}$ in the presence of the lattice
can be explained by the fact that it actually creates a nontrivial
\emph{band} (in terms of the corresponding values of $\mu $), where the solitons
may be stable (in the limit $\varepsilon \rightarrow 0$ the band degenerates
into a single point). Note that appearance of a similar band explains the
existence of stable 2D solitons in the OL-supported model with the \emph{
positive} scattering length \cite{baizakov}.
\begin{figure}[tbp]
\centerline{\scalebox{0.5}{\includegraphics[clip]{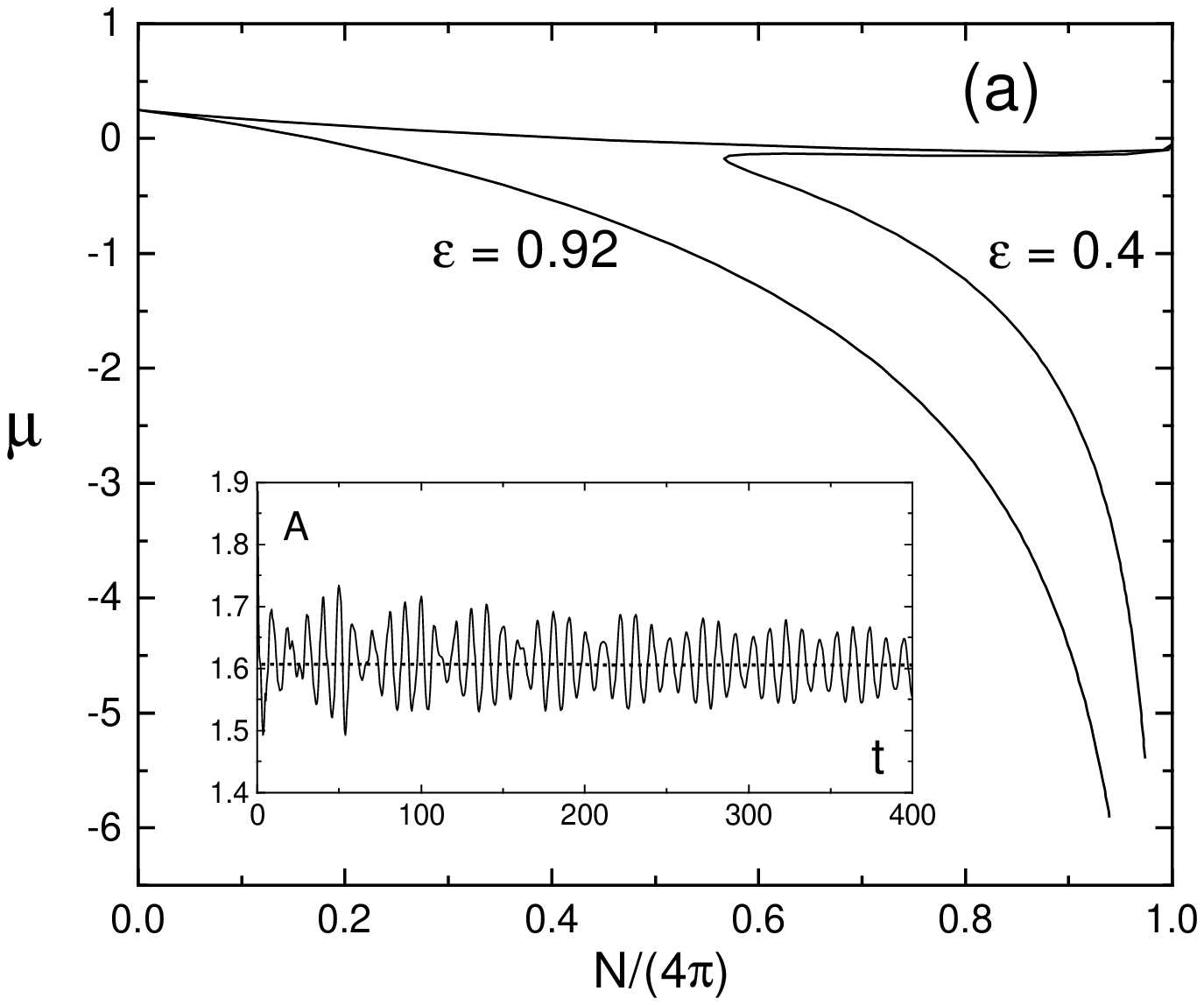}} \quad
            \scalebox{0.5}{\includegraphics[clip]{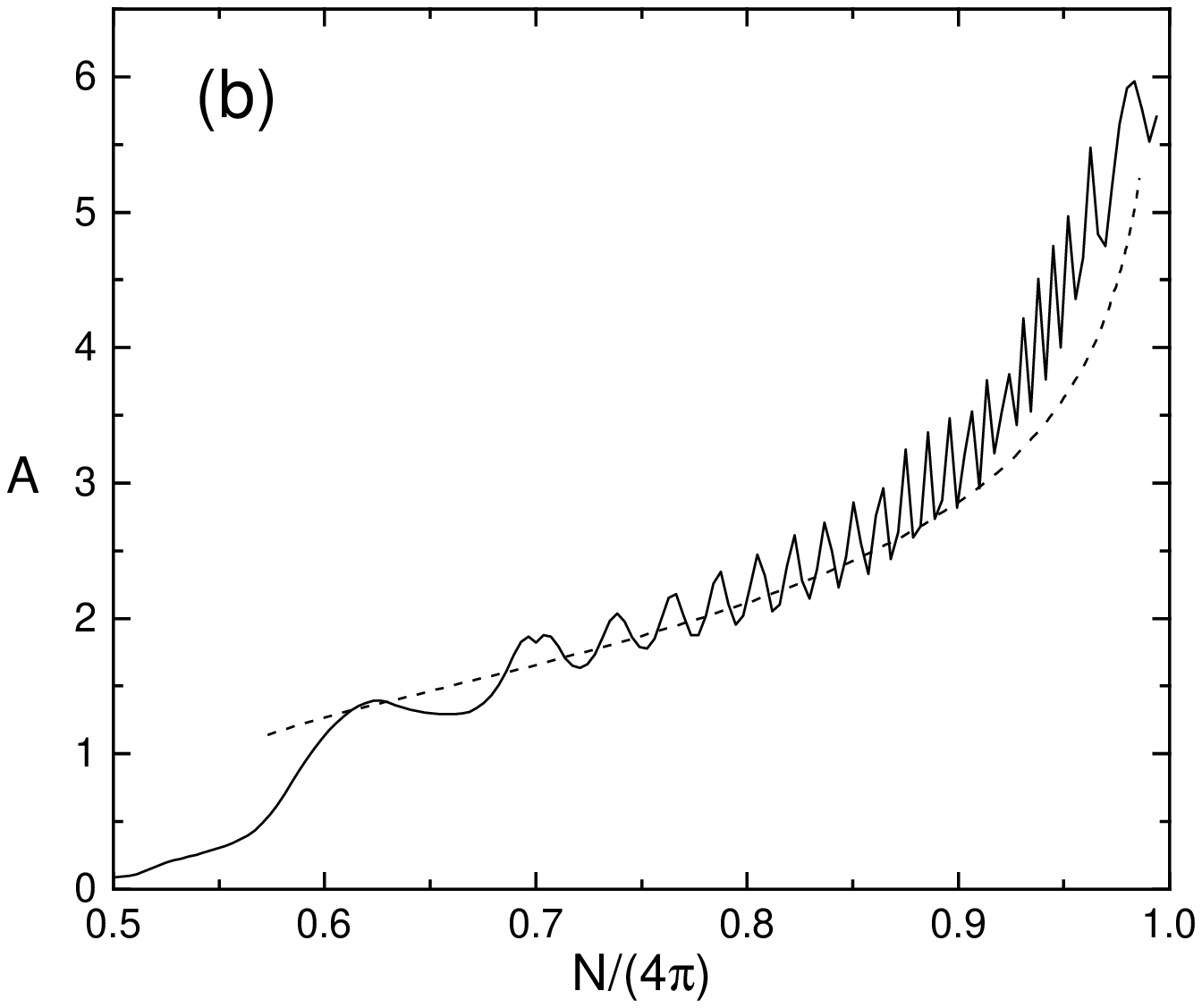}}}
\caption{(a) Numerical solution to Eqs. (\ref{2Dvar}) for
$\varepsilon =0.4$ and $\protect\varepsilon =\protect\varepsilon _{0}=0.92$
(the latter value is the one at which $N_{\mathrm{thr}}$ vanishes). Inset
displays evolution of the amplitude of a directly simulated solution
initiated by the variational ansatz (\ref{ansatz}) with $A$ and $a$ taken as
solutions of Eqs. (\ref{2Dvar}), in the case of $\protect\varepsilon =0.92$,
$N=2\protect\pi $, $a=1.3$ (the VK-stable branch). (b) Dependence of the
amplitude $A$ of established 2D solitons vs. the initial norm $N$, as
obtained from direct simulations of Eq. (\ref{gpe}) with
$\varepsilon =0.4$ and $\protect\omega =0$, starting with the configuration
predicted by VA. The undulations in the dependence is a result of radiation
loss in the course of the soliton formation. The dashed curve is the
dependence $A(N)$ as given by VA.}
\label{fig1}
\end{figure}

The simple ansatz (\ref{ansatz}) does not take into regard distortion of the
soliton's shape by the periodic potential; for this reason, VA is expected
to be accurate if the soliton's size is not essentially larger than the
lattice period, which is confirmed by comparison with numerical findings,
see below.

Direct simulations of Eq. (\ref{gpe}) were performed by means of the
multidimensional fast Fourier transform on grids of the size 256$\times $256
or 128$\times $128$\times $128 in the 2D and 3D cases, respectively, with a
time step $\Delta t=0.001$. The domain size was $\left| x,y\right| \leq 4\pi
$ (in the figures we show only its central part). To emulate infinite
boundary conditions (if the parabolic trap is switched off), absorbers were
installed at borders of the integration domain, which eliminate linear waves
emitted in the course of formation of solitons.

A basic conclusion following from the simulations is that the
initial waveforms, taken as predicted by VA, \emph{always} evolve
into stable solitons (in both 2D and 3D cases). On the other hand,
any input with the norm below the threshold predicted by VA (if
$\varepsilon <\varepsilon _{0}$), and some other input forms,
taken very far from the variational prediction, decay into linear
waves. This general result is illustrated by Fig. \ref{fig1}(b)
from which we see that a delocalizing transition, manifested by
the rapid drop of the amplitude, occurs at $N_{{\rm cr}}/4\pi
\approx 0.62$ for the case $\varepsilon =0.4$. By comparing this
value with the minimum value attained by $N/4\pi $ along the VA
existence curve of Fig. \ref{fig1}(a) at $\varepsilon =0.4$ (i.e.
$N_{min}/4\pi \approx 0.58$), we see that, although not perfect,
the agreement is reasonable. Similar results are found also
for other values of the amplitude of the optical lattice. We
remark, however, that non-sharpness of the transition
(see Fig. \ref{fig1}(b)) makes it difficult to determine the
threshold with higher accuracy since, close to the threshold, long
simulations are necessary to check whether a broad small-amplitude
soliton is formed or not. Accurate study of the threshold norm
versus $\varepsilon $ is presently under investigation and will be
reported elsewhere.

The presence of a lower threshold for the existence of the multi-dimensional
solitons in the OL strongly resemble a similar phenomenon occurring
in nonlinear lattices for intrinsic localized modes
\cite{kalosakas}. This analogy correlates with the fact that the above
solitons can be strongly localized around one site of the
lattice (see below).

Actually, the lattice stabilizes the soliton not only against decay,
but also partly against collapse. Indeed, in Fig. \ref{fig1}(b) the norm
extends to slightly overcritical values (the exact value in 2D is $N_{{\rm cr}}
/4\pi =0.93$ \cite{Berge}) without provoking the collapse.
An explanation for this comes from the
fact that the chemical potential of the soliton lies in the gap of the
excitation spectrum created by the lattice, thus enhancing their
stability against the collapse or decay into linear waves. The mechanism for
the formation of gap solitons in BECs with the OL and
{\em positive} scattering length was identified as
modulational instability of the Bloch states at the edges of the
Brillouin zone \cite{baizakov,KS02}.

Depending on the strength of the lattice
potential and norm of the initial pulse, the emerging soliton occupies a
single lattice cell (then we call it a ``single-cell soliton''), or spreads
itself over a few cells (a ``multi-cell soliton''). Single-cell solitons
always have a larger norm than the multi-cell ones, and a border value of
$N$, which separates the single- and multi-cell states, can be approximately
identified. The relaxation of the 2D and 3D pulses into the stable shape
proceeds with oscillations [see inset to Fig. \ref{fig1}(a)], with a higher
frequency in stronger lattices. In fact, the input which is predicted by VA
to be a stable soliton (according to the VK criterion) quickly evolves into
a single-cell soliton whose shape is quite close to the predicted one, which
is illustrated by inset to Fig. 1a. In contrast to this, the initial pulse
taken as what is expected to be a VK-unstable soliton, according to VA,
undergoes violent evolution, with large emission loss, which ends up by the
formation of a multi-cell soliton. Full evolution pictures are not displayed
here, as it is difficult to generate them in the 2D and 3D geometry without
radial symmetry.

\begin{figure}[tbp]
\centerline{\scalebox{0.4}{\includegraphics[clip]{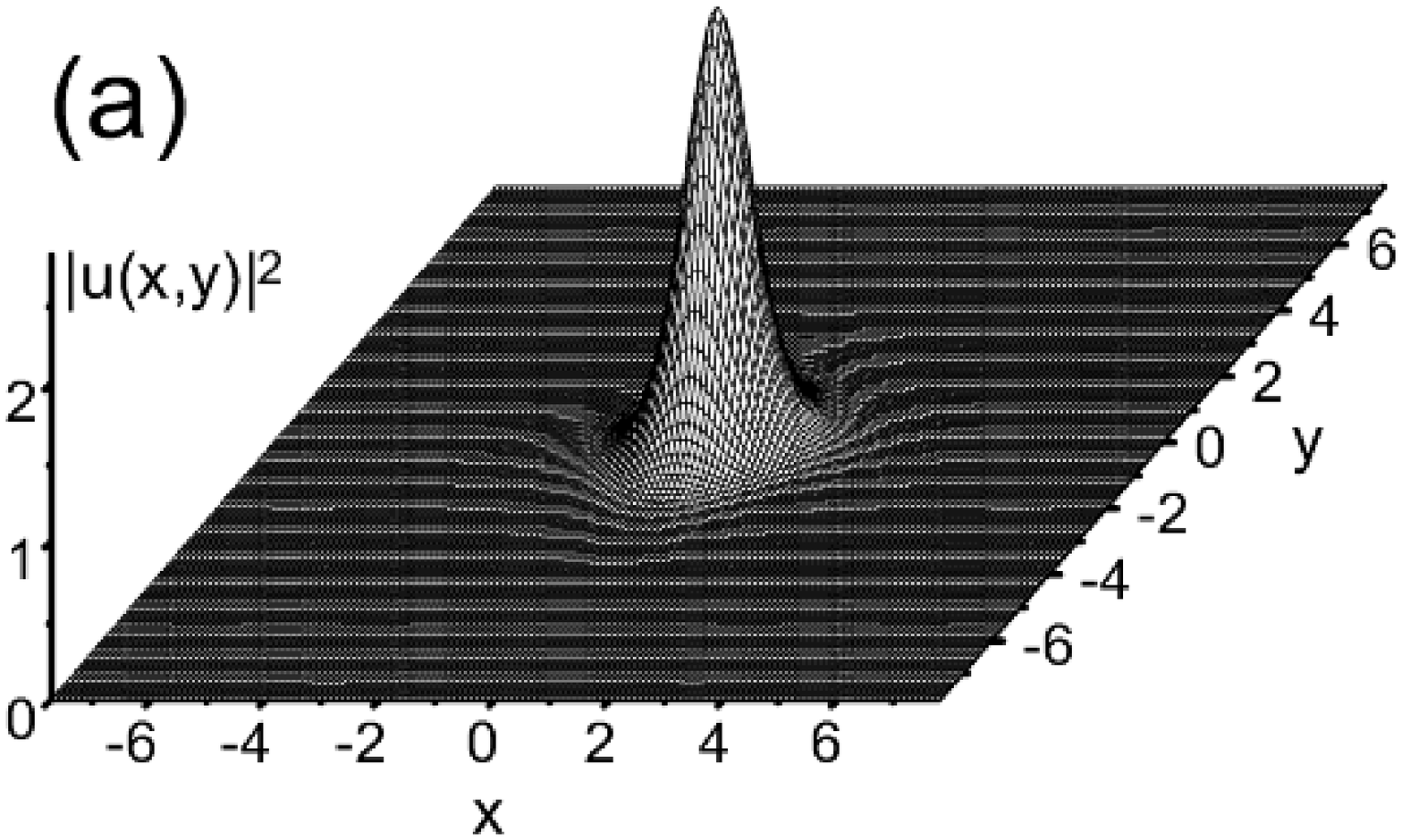}} \quad
            \scalebox{0.4}{\includegraphics[clip]{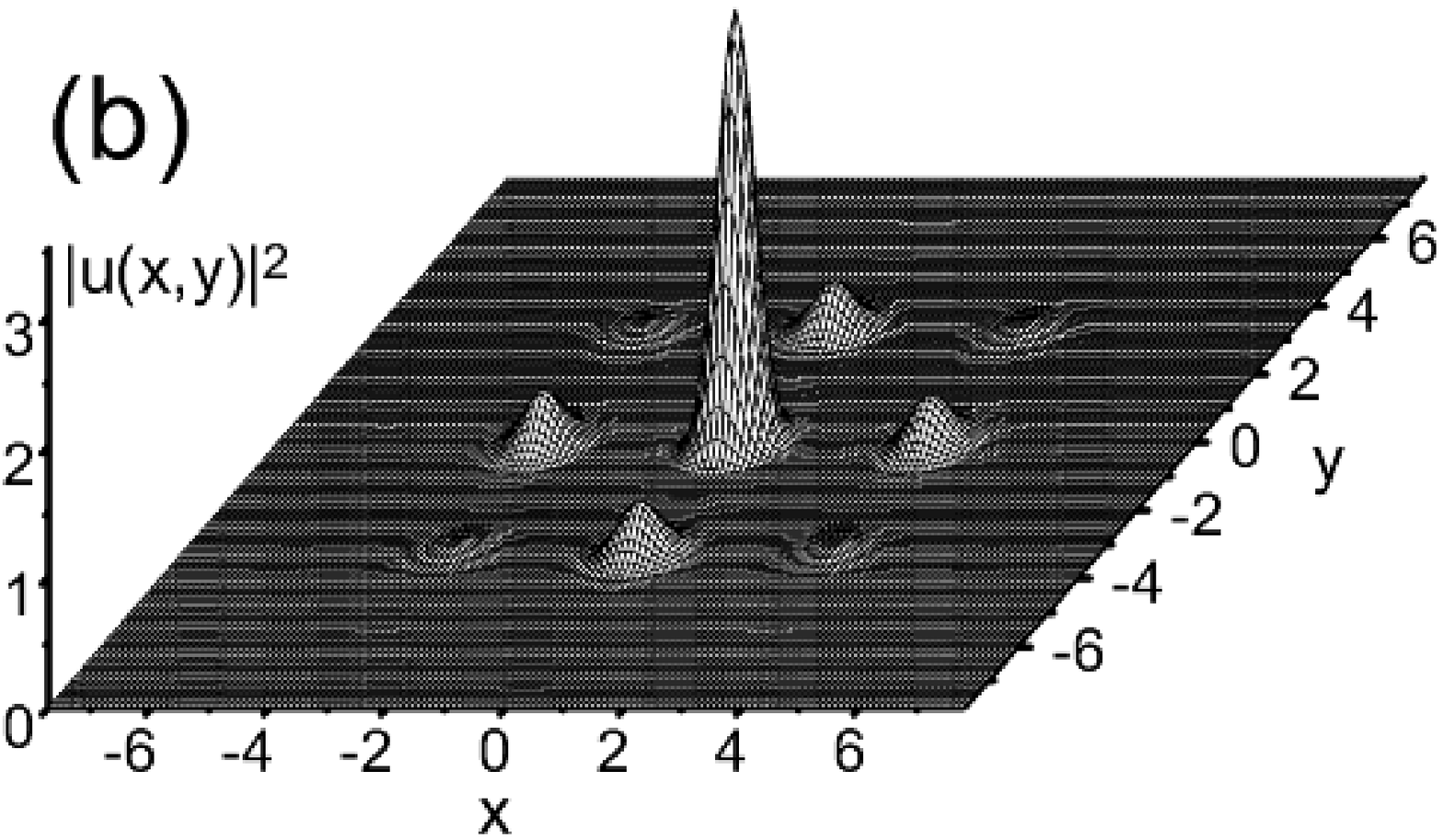}}}
\caption{(a) An established single-cell 2D soliton in a moderately strong
lattice potential, found from direct simulations of Eq. (\ref{gpe}) with
$\protect\varepsilon =0.92$. The initial configuration was taken as per the
VA-predicted stable soliton, i.e., Eqs. (\ref{ansatz}) and (\ref{2Dvar})
were used, with $a=1.3$, the corresponding norm being $N=2\protect\pi $. (b)
An established multi-cell 2D soliton in a strong lattice potential, with
$\protect\varepsilon =10$. The initial configuration was taken as per the
VA-predicted \emph{unstable} soliton with $a=0.15$ and $N=2\protect\pi$
[the same initial norm as in the case shown in (a)]. In the case (b), the
norm drops to $N=3.18$, i.e., $\approx 50\%$ of the initial norm is
lost in the course of the formation of the stable soliton.}
\label{fig2}
\end{figure}

Generic examples of the established single-cell and multi-cell 2D solitons,
which were generated, respectively, by the VA-predicted VK-stable Gaussian
ansatz in a moderately weak lattice potential, and the VK-unstable ansatz in
a strong potential, are displayed in Fig. \ref{fig2}. A central peak of the
soliton and satellites are clearly seen in the latter case.

\begin{figure}[tbp]
\centerline{\scalebox{0.5}{\includegraphics[clip]{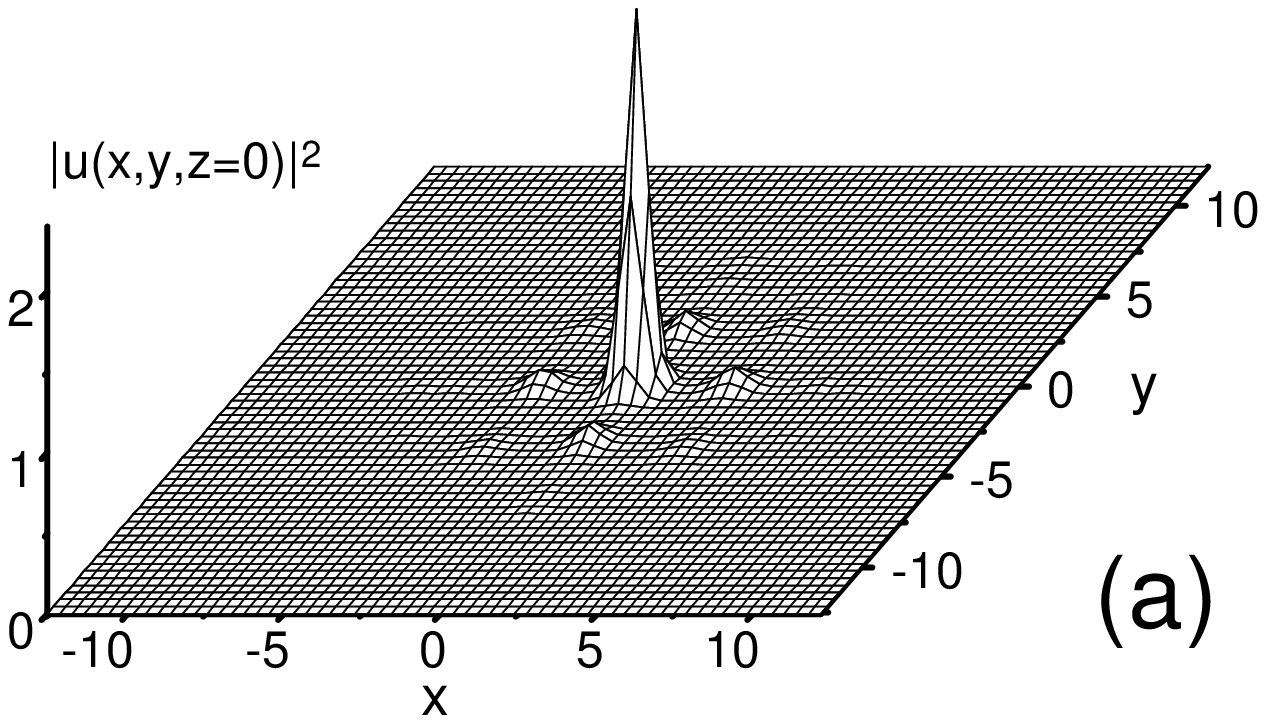}} \quad
            \scalebox{0.4}{\includegraphics[clip]{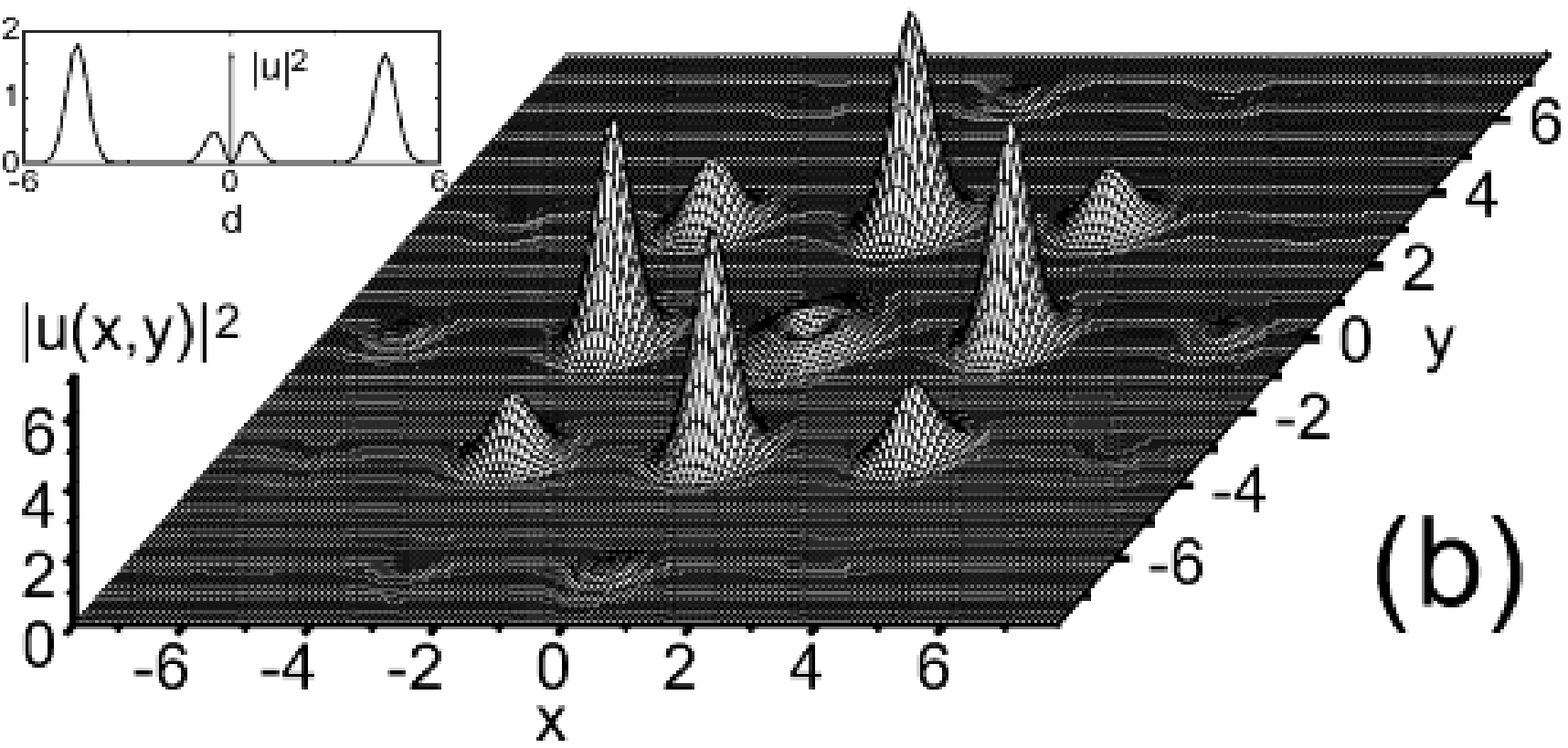}}}
\caption{(a) An established 3D soliton formed in a strong lattice potential.
The $z=0$ cross-section is shown for $\protect\varepsilon =10$ and $N=10$. (b)
A typical example of a stable vortex soliton with $S=1$ and $N=2\protect\pi$,
in the 2D lattice potential with $\protect\varepsilon =10$. Inset shows
the diagonal cross-section of the soliton; note that (as it should be) the
field vanishes $\sim r$ as $r\rightarrow 0$.}
\label{fig3}
\end{figure}

The formation of solitons in the 3D lattice potential is,
generally, similar to the 2D case, although the relaxation time is
shorter, due to the stronger interaction involving a larger number
of adjacent cells. A typical example of a \emph{stable} multi-cell
3D soliton, which forms itself in a strong lattice potential, is
given in Fig. \ref{fig3}(a).

Alongside the fundamental 2D soliton, in the
absence of the lattice one can construct vortex solitons, of the
form $u=v(r)\exp \left( -i\mu t+iS\theta \right) $, where $r$ and
$\theta $ are the polar coordinates, $S$ is an integer vorticity
(``spin''), and $v(r)$ is a real function, which exponentially
decays as $r\rightarrow \infty $ and vanishes $\sim r^{S}$ as
$r\rightarrow 0$. As well as the fundamental ($S=0$) soliton, the
vortices are strongly unstable in the usual 2D NLS equation.
Recently, stable vortices have been found in 2D (see reviews
\cite{Pramana}) and 3D \cite {Dumitru} models with non-Kerr
optical nonlinearities. On the other hand, it was demonstrated
\cite{Panos} that stable vortices, solely with $S=1$, exist in the
discrete version of the usual cubic 2D NLS equation (which
describes a bundle of nonlinear optical waveguides \cite{Moti}, or
BEC trapped in a strong OL field \cite{Smerzi}. Note that, unlike
the isotropic NLS model, in ones with the axial symmetry broken by
the lattice the vorticity is not a topological invariant, hence
the very existence of such solutions is a nontrivial issue
\cite{Panos}.

To generate vortex solitons, we simulated the 2D version of Eq. (\ref{gpe})
with the initial condition $u_{0}(r,\theta )=Ar^{S}\exp (-ar^{2})\exp
(iS\theta )$, trying various values of $A$ and $a$. Stable vortices with
$S\geq 2$ could never be created (they quickly spread out); however, initial
configurations with $S=1$ readily self-trap into a stable vortex, which
features a complex multi-cell organization, see Fig. \ref{fig3}(b).
All the above examples were obtained for $\omega =0$ in Eq. (\ref{gpe}),
i.e., without the parabolic trap. Simulations with $\omega \neq 0$ have
demonstrated that the external trap does not affect the 2D and 3D solitons
in any conspicuous way, provided that the corresponding harmonic-oscillator
length is essentially larger than the lattice period.

The solitons found in this work are pinned by the lattice. In fact, a
similar mechanism may stabilize solitons that partially keep
the mobility of their free counterparts. To this end, one can use quasi-1D
and quasi-2D lattices, in the 2D and 3D cases, respectively. This
modification of the model will be reported elsewhere.

Finally, we address experimental perspectives. First, creation of the
initial waveform occupying a single lattice cell can be done
using the recently developed technique for patterned loading of BEC
into optical lattices \cite{peil}, which provides flexible control over
placement of atoms in lattice sites. An initial waveform spread over
multiple cells can be prepared by imposing an OL upon a condensate of a
suitable size in the magnetic trap, with subsequent switching off the
magnetic field.
Changing the norm of the wave function, as supposed in the
theory and numerical simulations, could be modeled by variation of the
nonlinear coupling parameter (s-wave scattering length) via the Feshbach
resonance.

We note that the strengths of the OLs considered above are in the
experimentally relevant range \cite{OLlater}, $\varepsilon =0\div 20$ in
units of the recoil energy $E_{\mathrm{rec}}=\hbar ^{2}k_{L}^{2}/(2m)$,
where $k_{L}=2\pi /\lambda $, and $m$ and$\ \lambda $ denote the atomic mass
and laser wavelength, respectively. We used the size of a unit cell $d\equiv
\lambda /2=0.425$ \thinspace\ $\mu $m, and the time $2m/(\hbar k_{L}^{2})=50$
\thinspace\ $\mu $s (for $^{85}$Rb atoms and far detuned laser with $\lambda
=850$ nm) as units of the length and time, respectively. Relevant
experiments with can be performed with $^{7}$Li or $^{85}$Rb atoms featuring
the negative scattering length in the ground state, which is amenable to
large variations through the Feshbach resonance.

In conclusion, we have demonstrated that the GP/NLS equation with the
attractive cubic nonlinearity and periodic potential, which describes BECs
trapped in a 2D or 3D optical lattice, and (in the 2D case) an optical beam
in the Kerr medium with a transverse periodic modulation of the refractive
index, gives rise to stable solitons. In moderately weak and strong
lattices, single-cell and multi-cell solitons were found, respectively, the
former ones being accurately predicted by the variational approximation. A
necessary condition for the formation of 2D solitons in a relatively weak
lattice is that the initial norm of the field must exceed a threshold value.
Stable 2D vortex solitons with $S=1$ were found too.

\acknowledgments
B.B.B. thanks the Physics Department of the University of Salerno, Italy,
for a two-years research grant. B.A.M. appreciates hospitality of the same
Department and thanks P.G. Kevrekidis for valuable discussions.
M.S. acknowledges partial financial support from the MIUR, through the
inter-university project PRIN-2000, and from the European grant
LOCNET no. HPRN-CT-1999-00163.

\end{document}